\documentclass[twocolumn,english,prb,aps,superscriptaddress]{revtex4}
\usepackage{lmodern}

\usepackage[latin9]{inputenc}
\usepackage{amsmath}
\usepackage{amssymb}
\usepackage{graphicx}
\usepackage{esint}

\makeatletter
\@ifundefined{textcolor}{}
{%
 \definecolor{BLACK}{gray}{0}
 \definecolor{WHITE}{gray}{1}
 \definecolor{RED}{rgb}{1,0,0}
 \definecolor{GREEN}{rgb}{0,1,0}
 \definecolor{BLUE}{rgb}{0,0,1}
 \definecolor{CYAN}{cmyk}{1,0,0,0}
 \definecolor{MAGENTA}{cmyk}{0,1,0,0}
 \definecolor{YELLOW}{cmyk}{0,0,1,0}
}

\usepackage[USenglish]{babel}
\@ifundefined{definecolor}{\usepackage{color}}{}
\DeclareMathOperator{\sgn}{sgn}
\newcommand{\beq}{\begin{equation}}
\newcommand{\eeq}{\end{equation}}
\newcommand{\bea}{\begin{eqnarray}}
\newcommand{\eea}{\end{eqnarray}}
\newcommand{\be}{\begin{equation}}
\newcommand{\ee}{\end{equation}}

\renewcommand{\phi}{\varphi}

\usepackage{babel}

\makeatother

\usepackage{babel}
\begin{document}

\title{Time-reversal symmetry breaking superconductivity in the coexistence
phase with magnetism in Fe-pnictides}

\author{Alberto Hinojosa}

\affiliation{Department of Physics, University of Wisconsin, Madison, WI 53706,
USA}

\author{Rafael M. Fernandes}

\affiliation{School of Physics and Astronomy, University of Minnesota, Minneapolis,
MN 55455, USA}

\author{Andrey V. Chubukov}

\affiliation{Department of Physics, University of Wisconsin, Madison, WI 53706,
USA}

\date{\today}
\begin{abstract}
We argue that superconductivity in the coexistence region with spin-density-wave
(SDW) order in weakly doped Fe-pnictides differs qualitatively from
the ordinary $s^{+-}$ state outside the coexistence region, as it
develops an additional gap component which is a mixture of intra-pocket
singlet ($s^{++}$) and inter-pocket spin-triplet pairings (the $t-$state).
The coupling constant for the $t-$channel is proportional to the
SDW order and involves interactions that do not contribute to superconductivity
outside of the SDW region. We argue that the $s^{+-}$ and $t-$type
superconducting orders coexist at low temperatures, and the relative
phase between the two is in general different than $0$ or $\pi$,
manifesting explicitly the breaking of the time-reversal symmetry
promoted by long-range SDW order. We show that this exotic state emerges
already in the simplest model of Fe-pnictides, with one hole pocket
and two symmetry-related electron pockets. We argue that in some parameter
range time-reversal gets broken even before long-range superconducting
order develops.
\end{abstract}
\maketitle
\textit{Introduction}~~~Iron-based superconductors (FeSCs) have
been the subject of intense study since 2008 \cite{reviews}. Their
rich phase diagram includes the regions of superconductivity (SC),
spin density wave (SDW), nematic order, and a region where SDW, SC,
and nematic order coexist \cite{review_nematics}. Outside the SDW/nematic
region, SC develops in the spin-singlet channel and in most of Fe-based
superconductors has $s-$wave symmetry with a $\pi$ phase shift between
the SC order parameters on hole and on electron pockets ( $s^{+-}$
gap structure) \cite{magnetic,reviews_pairing}.

It has been recently argued by several groups that the multiband structure
of FeSCs allows for superconducting states with more exotic properties
\cite{CWu09,Stanev10,Babaev11,Maiti12,s_plus_id_Maier,s_plus_id_Thomale,s_plus_id_Khodas,Livanas12,Benfatto13,DHLee13,Kotliar13,Hu14,Fernandes13,s_plus_is,Chubukov_LiFeAs}.
Of particular interest are SC states that break time-reversal symmetry
(TRS), as such states have a plethora of interesting properties like,
e.g., novel collective modes~\cite{s_plus_is,Benfatto13,Stanev12,Lin12}.
TRS-broken states emerge when the phase differences $\psi_{i}$ between
SC order parameters on different Fermi surfaces (FS) are not multiples
of $\pi$.

The two current proposals for TRS breaking in FeSCs are $s+id$ \cite{CWu09,s_plus_id_Maier,s_plus_id_Thomale,s_plus_id_Khodas,Fernandes13}
and $s+is$ states \cite{Stanev10,s_plus_is,Benfatto13,Chubukov_LiFeAs}.
The first emerges when attractions in the $d-$wave and $s-$wave
channels are of near-equal strength. The second emerges when there
is a competition between different $s^{+-}$ states favored by inter-pocket
and intra-pocket interactions. Both of these proposals were, however,
argued to be applicable only to strongly hole or electron-doped FeSCc.
For weakly/moderately doped FeSCs the common belief is that $s^{+-}$
superconductivity is robust.

In this communication we argue that an exotic state which breaks TRS
can emerge already at low doping, in a range where SC is known \cite{coexist_1,coexist_2,coexist_3,coexist_4,coexist_5,coexist_6,coexist_7,coexist_8,coexist_9}
to emerge from a pre-existing SDW state. Previous works on SC in the
coexistence region focused on the SDW-induced modification of the
form of $s^{+-}$ gap \cite{parker_09,vvc,fs,Eremin11,Maiti12_nodes,Dagotto14,Timm14}.
We argue that there is another effect -- SDW order also induces attraction
in another pairing channel, for which the order parameter is an admixture
of spin-singlet and spin-triplet components (the two are mixed in
the SDW state since spin rotational symmetry is broken). Because a
triplet component is involved, we will be calling this state as $t$-state.
The coupling in the $t-$channel is a combination of interactions
that do not contribute to $s^{+-}$ SC in the paramagnetic state.
A real admixture between these singlet and triplet SC states, $s\pm t$,
has been discussed in the SDW/SC coexistence region of organics, cuprates,
and heavy fermions \cite{triplet_organics,triplet_Fukuyama,triplet_Kyung,triplet_heavy_fermions,triplet_Eremin,triplet_GL,triplet_cuprates}.
Here, however, we found that the situation is different -- $s\pm t$
state exists only near $T_{c}$, while at low $T$, the relative phase
between the two SC components is different from $0$ or $\pi$, i.e.,
the order parameter has $s+\mathrm{e}^{i\theta}t$ form. This order
parameter does not transform into itself under TRS, unlike $s\pm t$
order. As a result, the order parameter manifold contains an additional
$Z_{2}$ Ising degree of freedom, which gets broken by selection of
$+\theta$ or $-\theta$. The TRS broken state emerges via a phase
transition inside a superconductor, which should have experimental
manifestations. We note in this regard that that, although TRS of
the system is formally broken already at the SDW transition temperature
$T_{N}>T_{c}$, the TR operation transforms one magnetic state into
another state from the same $O(3)$ manifold, i.e., there is no additional
$Z_{2}$ degree of freedom which one could associate with TRS. The
$s^{+-}$ state also does not contain this extra degree of freedom
simply because it transforms into itself under TRS. Only when $\theta$
becomes different from $0$ or $\pi$, does the order parameter manifold
acquire an additional $Z_{2}$ degree of freedom associated with TRS.

We show that the $s+\mathrm{e}^{i\theta}t$ state emerges already in the
simplest three-band model of one circular hole pocket and two symmetry-related
elliptical electron pockets \cite{Eremin10}. Since SDW order in most
of the range where SC and SDW coexist is of stripe type, the associated
FS reconstruction involves only one hole and one electron pocket separated
by either $(0,\pi)$ or $(\pi,0)$ in the 1-Fe Brillouin zone, reducing
the model to a two-pocket model \cite{vvc,fs}.

The pairing interaction in the $t-$channel emerges once the original
4-fermion interactions for the two pockets connected by the SDW ordering
vector are dressed up by SDW coherent factors. When the pairing interactions
are rewritten in terms of $a$ and $b$ fermions, which describe states
near the reconstructed FSs, they yield conventional terms like $a_{\mathbf{k}\uparrow}^{\dagger}a_{-\mathbf{k}\downarrow}^{\dagger}a_{-\mathbf{p}\downarrow}a_{\mathbf{p}\uparrow}$
or $a_{\mathbf{k}\uparrow}^{\dagger}a_{-\mathbf{k}\downarrow}^{\dagger}b_{-\mathbf{p}\downarrow}b_{\mathbf{p}\uparrow}$,
and also anomalous terms like $a_{\mathbf{k}\uparrow}^{\dagger}a_{-\mathbf{k}\downarrow}^{\dagger}a_{-\mathbf{p}\downarrow}b_{\mathbf{p}\uparrow}$.
As a consequence, spin-singlet pairing between FSs of the same kind
($i\sigma_{\alpha\beta}^{y}\langle a_{\mathbf{k}\alpha}a_{-\mathbf{k}\beta}\rangle$
and $i\sigma_{\alpha\beta}^{y}\langle b_{\mathbf{k}\alpha}b_{-\mathbf{k}\beta}\rangle$)
mixes with spin triplet pairing between FSs of opposite type ($\sigma_{\alpha\beta}^{x}\langle a_{\mathbf{k}\alpha}b_{-\mathbf{k}\beta}\rangle$).
We show below that this gives rise to the emergence of two different
superconducting channels. One is the usual spin-singlet $s^{+-}$
channel, for which the SC order parameter is $\Delta_{1}\propto\sum_{\mathbf{k}}i\sigma_{\alpha\beta}^{y}\left[\langle a_{\mathbf{k}\alpha}a_{-\mathbf{k}\beta}\rangle-\langle b_{\mathbf{k}\alpha}b_{-\mathbf{k}\beta}\rangle\right]$.
If only this SC develops, the gaps on the two FSs have a phase difference
of $\pi$ (we define SC order parameters such that in the absence
of SDW $\langle a_{\mathbf{k}\alpha}a_{-\mathbf{k}\beta}\rangle$
and $\langle b_{\mathbf{k}\alpha}b_{-\mathbf{k}\beta}\rangle$ become
the SC order parameters on the hole and electron FSs). The second
pairing channel,
with order parameter $\Delta_{2}$, has two contributions. One is
a spin-triplet inter-pocket term $\sum_{\mathbf{k}}\sigma_{\alpha\beta}^{x}\langle a_{\mathbf{k}\alpha}b_{-\mathbf{k}\beta}\rangle$)
(hence the name $t-$state), and the other is a spin-singlet $s^{++}$
type term $\sum_{\mathbf{k}}i\sigma_{\alpha\beta}^{y}\left[\langle a_{\mathbf{k}\alpha}a_{-\mathbf{k}\beta}\rangle+\langle b_{\mathbf{k}\alpha}b_{-\mathbf{k}\beta}\rangle\right]$.
The presence of the $s^{++}$ component in $\Delta_{2}$ is crucial
as with it the kernel in the gap equation for $\Delta_{2}$ is logarithmical
(as it is for $\Delta_{1}$), implying that even a weak attraction
in this channel gives rise to superconductivity. A similar situation
emerges in Fe-pnictides with only electron pockets -- the analog of
$\langle ab \rangle$ term there is induced by hybridization~\cite{Khodas2012}.

\begin{figure}[htb]
\centering \includegraphics[width=0.35\textwidth]{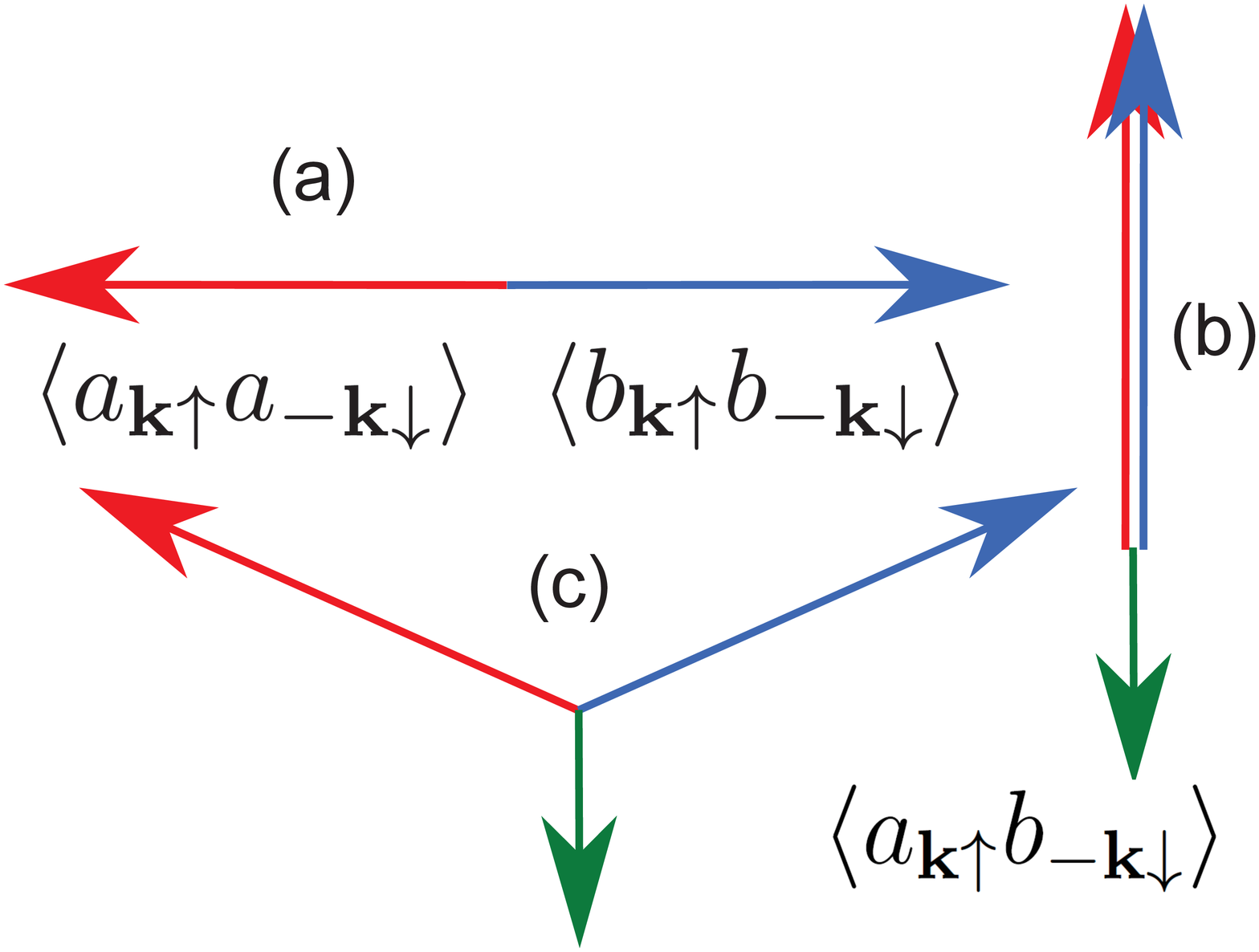} \caption{The structure of gap functions in different SC states: (a) pure $s^{+-}$
state, (b) pure $t-$ state, (c) $s+it$ state with $\pm\pi/2$ phase
difference between the phases of $s^{+-}$ and $t-$ gaps. Operators
$a$ and $b$ describe fermions near the reconstructed FSs. \label{fig:phasors}}
\end{figure}

The structure of $\Delta_{1}$ and $\Delta_{2}$ is shown in Figs.
\ref{fig:phasors}a and \ref{fig:phasors}b. Our analysis of the non-linear
gap equations for $\Delta_{1}$ and $\Delta_{2}$ shows that the two
SC orders coexist in some parameter range, and the relative phase
between the two is different than $0$ or $\pi$, in the general case
when the two orders are linearly coupled in the Ginzburg-Landau (GL)
functional, and equals to $\pm\pi/2$ for the special case when linear
coupling is absent (Fig. \ref{fig:phasors}c).

\begin{figure}
\includegraphics[width=0.45\textwidth]{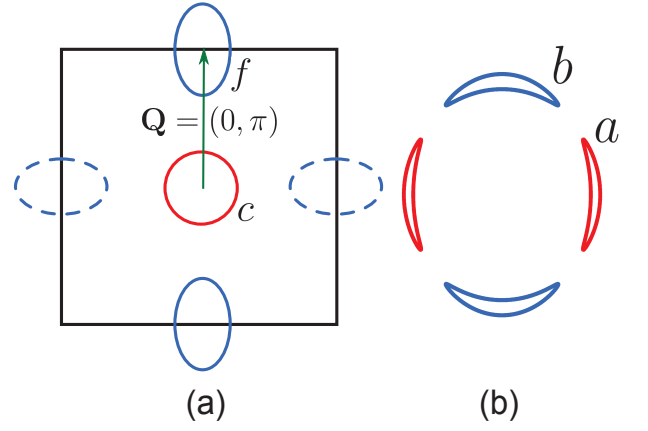} \caption{Fermi surfaces in (a) the paramagnetic state, (b) the SDW state.\label{fig:Normal-SDW_FS}}
\end{figure}

\textit{The model.}~~~ We consider a three band model with $c$
fermions with momenta near the hole pocket at $(0,0)$ and $f$ fermions
with momenta near the electron pockets centered at $(0,\pi)$ and
$(\pi,0)$
 in the 1-Fe Brillouin zone (Fig. \ref{fig:Normal-SDW_FS}a) \cite{Eremin10,Fernandes12}.
The $c$ and $f$ fermions form circular and elliptical FSs, respectively,
with dispersions given by $\xi_{\mathbf{k}}^{c}=\mu_{c}-\frac{\mathbf{k}^{2}}{2m_{c}}$
and $\xi_{\mathbf{k}}^{f}=-\mu_{f}+\frac{k_{x}^{2}}{2m_{x}}+\frac{k_{y}^{2}}{2m_{y}}$.
Since the SDW state picks an ordering vector $\mathbf{Q}$, which
is either $(0,\pi)$ or $(\pi,0)$, one of the electron pockets does
not participate in this order. We choose $\mathbf{Q}=(0,\pi)$ without
loss of generality and effectively reduce the model to two bands.
We follow earlier works~\cite{Chubukov_PhysicaC,Maiti10} and consider
five possible repulsive interactions in the band basis:
inter-pocket, density-density, exchange, pair hopping, and intra-pocket
interactions. The corresponding couplings are $U_{1},U_{2},U_{3}$,
and $U_{4}$, respectively. We present the interaction Hamiltonian
in the Supplementary material (SM). All couplings are assumed to be
already renormalized from their bare values by fermions with energies
larger than the upper energy cutoff $\Lambda$. Without SDW, SC in
this model arises only in the $s^{+-}$ channel. The corresponding
coupling is $U_{3}-U_{4}$, and we assume that it is positive (attractive).
The couplings $U_{1}$ and $U_{2}$ do not participate in SC pairing,
but $U_{1}$ contributes to the coupling in the SDW channel $U_{1}+U_{3}>0$,
which for $U_{i}>0$ is larger than in SC channels, i.e., the system
first develops SDW order upon lowering $T$, and superconductivity
emerges from a pre-existing SDW state. RG studies 
found that the SC interaction gets larger as energy decreases in the
RG flow \cite{Chubukov_PhysicaC,Maiti10,thomale,dhl}. Yet, at low dopings, the
SDW order comes first and SC develops in the coexistence
region with magnetism.

The self-consistent equation for the SDW order parameter $M$ and
the reconstructed fermionic dispersions in the SDW state have been
obtained before \cite{Eremin10}. The quadratic Hamiltonian in terms
of the new quasiparticles $a$ and $b$ is
\begin{equation}
\mathcal{H}_{0}=\sum_{\mathbf{k}}\left[\xi_{\mathbf{k}}^{a}a_{\mathbf{k}\alpha}^{\dagger}a_{\mathbf{k}\alpha}+\xi_{\mathbf{k}}^{b}b_{\mathbf{k}\alpha}^{\dagger}b_{\mathbf{k}\alpha}\right],
\end{equation}
 where
\begin{align}
\xi_{\mathbf{k}}^{a} & =\delta_{\mathbf{k}}-\sqrt{\xi_{\mathbf{k}}^{2}+M^{2}},\\
\xi_{\mathbf{k}}^{b} & =\delta_{\mathbf{k}}+\sqrt{\xi_{\mathbf{k}}^{2}+M^{2}},
\end{align}
 and we have expressed the original dispersions in terms of the linear
combinations $\delta_{\mathbf{k}}=\frac{\xi_{\mathbf{k}}^{f}+\xi_{\mathbf{k}}^{c}}{2}$
and $\xi_{\mathbf{k}}=\frac{\xi_{\mathbf{k}}^{f}-\xi_{\mathbf{k}}^{c}}{2}$.
In general $\delta_{\mathbf{k}}=\delta_{0}+\delta_{2}\cos{2\theta}$,
where the first term measures the doping ($\delta_{0}=0.5v_{F}(k_{F}^{c}-k_{F}^{f})$)
and the second one accounts for the (weak) ellipticity of the electron
pocket (Ref.~\cite{vvc}). The coherence factors $u_{\mathbf{k}}$
and $v_{\mathbf{k}}$ are expressed in terms of these parameters as
$u_{\mathbf{k}}=\sqrt{\frac{1}{2}\left(1+\frac{\xi_{\mathbf{k}}}{\sqrt{\xi_{\mathbf{k}}^{2}+M^{2}}}\right)}$,
$v_{\mathbf{k}}=\sgn M\sqrt{\frac{1}{2}\left(1-\frac{\xi_{\mathbf{k}}}{\sqrt{\xi_{\mathbf{k}}^{2}+M^{2}}}\right)}$ (see SM).
The FSs for $a$ and $b$ fermions are shown in Fig. \ref{fig:Normal-SDW_FS}b.

\textit{Superconductivity}.~~~ We now consider the pairing interactions
leading to SC inside the SDW state. As a first step, we rewrite the
interactions
 in terms of the new fermions.
  We then find conventional pairing terms like $a_{\mathbf{k}\uparrow}^{\dagger}a_{-\mathbf{k}\downarrow}^{\dagger}a_{-\mathbf{p}\downarrow}a_{\mathbf{p}\uparrow}$
or $a_{\mathbf{k}\uparrow}^{\dagger}a_{-\mathbf{k}\downarrow}^{\dagger}b_{-\mathbf{p}\downarrow}b_{\mathbf{p}\uparrow}$,
and anomalous terms like $a_{\mathbf{k}\uparrow}^{\dagger}a_{-\mathbf{k}\downarrow}^{\dagger}(a_{-\mathbf{p}\downarrow}b_{\mathbf{p}\uparrow}+a_{-\mathbf{p}\uparrow}b_{\mathbf{p}\downarrow})$.
To solve for the SC order parameter, we then need to introduce both
spin-singlet pairings $i\sigma_{\alpha\beta}^{y}\langle a_{\mathbf{k}\alpha}a_{-\mathbf{k}\beta}\rangle$
and $i\sigma_{\alpha\beta}^{y}\langle b_{\mathbf{k}\alpha}b_{-\mathbf{k}\beta}\rangle$
between fermions belonging to the same pocket, and spin triplet pairing
$\sigma_{\alpha\beta}^{x}\langle a_{\mathbf{k}\alpha}b_{-\mathbf{k}\beta}\rangle$
between fermions belonging to different pockets.

The full pairing Hamiltonian in the BCS approximation has the form
\begin{align}
\mathcal{H}_{\Delta} & =\frac{1}{2}\sum_{\mathbf{p}}\Delta_{aa}(\mathbf{p})i\sigma_{\alpha\beta}^{y}a_{\mathbf{p}\alpha}^{\dagger}a_{-\mathbf{p}\beta}^{\dagger}\label{eq:mfH}\\
 & +\frac{1}{2}\sum_{\mathbf{p}}\Delta_{bb}(\mathbf{p})i\sigma_{\alpha\beta}^{y}b_{\mathbf{p}\alpha}^{\dagger}b_{-\mathbf{p}\beta}^{\dagger}\nonumber \\
 & +\frac{1}{2}\sum_{\mathbf{p}}\Delta_{ab}(\mathbf{p})\sigma_{\alpha\beta}^{x}[a_{\mathbf{p}\alpha}^{\dagger}b_{-\mathbf{p}\beta}^{\dagger}-b_{\mathbf{p}\alpha}^{\dagger}a_{-\mathbf{p}\beta}^{\dagger}]+\mathrm{H.c.}\nonumber
\end{align}
 Because there are three different anomalous terms, the diagonalization
of the pairing Hamiltonian leads to a set of three coupled equations
for $\Delta_{aa}$, $\Delta_{bb}$, and $\Delta_{ab}$. Parameterizing
$\Delta_{ij}$ as
\begin{align}
\Delta_{aa,bb}(\mathbf{p}) & =\pm\Delta_{1}+\Delta_{2}(2u_{\mathbf{p}}v_{\mathbf{p}})+\Delta_{3}(u_{\mathbf{p}}^{2}-v_{\mathbf{p}}^{2}),\label{eq:gaps1}\\
\Delta_{ab}(\mathbf{p}) & =\Delta_{2}(u_{\mathbf{p}}^{2}-v_{\mathbf{p}}^{2})-\Delta_{3}(2u_{\mathbf{p}}v_{\mathbf{p}}),\label{eq:gaps2}
\end{align}
 we express the equations for SC order parameters as
\begin{align}
\Delta_{1} & =\frac{U_{3}-U_{4}}{2}\sum_{\mathbf{k}}\left[\langle aa\rangle_{\mathbf{k}}-\langle bb\rangle_{\mathbf{k}}\right],\label{eq:delta1}\\
\Delta_{2} & =(U_{2}-U_{1})\sum_{\mathbf{k}}\left[u_{\mathbf{k}}v_{\mathbf{k}}(\langle aa\rangle_{\mathbf{k}}+\langle bb\rangle_{\mathbf{k}})+(u_{\mathbf{k}}^{2}-v_{\mathbf{k}}^{2})\langle ab\rangle_{\mathbf{k}}\right],\\
\Delta_{3} & =-\frac{U_{3}+U_{4}}{2}\sum_{\mathbf{k}}\left[(u_{\mathbf{k}}^{2}-v_{\mathbf{k}}^{2})(\langle aa\rangle_{\mathbf{k}}+\langle bb\rangle_{\mathbf{k}})-4u_{\mathbf{k}}v_{\mathbf{k}}\langle ab\rangle_{\mathbf{k}}\right].\label{eq:delta3}
\end{align}
 where $\langle aa\rangle_{\mathbf{k}}\equiv i\sigma_{\alpha\beta}^{y}\langle a_{-\mathbf{k}\beta}a_{\mathbf{k}\alpha}\rangle$,
$\langle bb\rangle_{\mathbf{k}}\equiv i\sigma_{\alpha\beta}^{y}\langle b_{-\mathbf{k}\beta}b_{\mathbf{k}\alpha}\rangle$,
$\langle ab\rangle_{\mathbf{k}}\equiv\sigma_{\alpha\beta}^{x}\langle b_{-\mathbf{k}\beta}a_{\mathbf{k}\alpha}\rangle$.
Each average is in turn expressed in terms of $\Delta_{i}$ ($i=1,2,3)$,
i.e. Eqs. (\ref{eq:delta1})-(\ref{eq:delta3}) represent the set
of three coupled non-linear equations for the SC order parameters
in the presence of SDW order.

We see from (\ref{eq:delta1}) that three combinations of the interactions
$U_{i}$ appear in the pairing channel. Two have familiar forms \cite{Chubukov_PhysicaC}:
$U_{3}-U_{4}$ and $-(U_{3}+U_{4})$ are the couplings in the $s^{+-}$
and $s^{++}$ channels, respectively, in the absence of SDW order.
A non-zero $M$ couples the $s^{+-}$ and $s^{++}$ channels, but
since the coupling in the $s^{++}$ channel is strongly repulsive,
the SDW-induced mixing of $s^{+-}$ and $s^{++}$ channels should
not lead to any new physics. The third coupling $U_{2}-U_{1}$, on
the other hand, does not contribute to SC in the absence of SDW order.
Its presence in Eq. (\ref{eq:delta1}) implies that SDW order not
only modifies the two existing pairing channels, but also generates
a new channel of fermionic pairing.

We present the full expressions for $\langle ij\rangle_{\mathbf{k}}$
in the SM and here focus on the linearized gap equations, valid at
the corresponding $T_{c,i}$. Expanding
 the r.h.s. of (\ref{eq:delta1}) to first order in $\Delta_{ij}$
we obtain
\begin{align}
\langle aa\rangle_{\mathbf{k}}\pm\langle bb\rangle_{\mathbf{k}} & =\frac{\Delta_{aa}(k)}{2\xi_{k}^{a}}\tanh{\frac{\xi_{k}^{a}}{2T}}\pm\frac{\Delta_{bb}(k)}{2\xi_{k}^{b}}\tanh{\frac{\xi_{k}^{b}}{2T}}\nonumber \\
\langle ab\rangle_{\mathbf{k}} & =\frac{\Delta_{ab}(k)}{2(\xi_{k}^{a}+\xi_{k}^{b})}\left(\tanh{\frac{\xi_{k}^{a}}{2T}}+\tanh{\frac{\xi_{k}^{b}}{2T}}\right)\label{ch1}
\end{align}
 where $\Delta_{ij}$ are expressed via $\Delta_{i}$ by Eq. (\ref{eq:gaps2}).
Substituting (\ref{ch1}) into the r.h.s. of (\ref{eq:delta1}) we
obtain the set of three coupled linearized Eqs. on $\Delta_{i}$ which
can be easily solved.

\begin{figure}
\centering \includegraphics[width=0.45\textwidth]{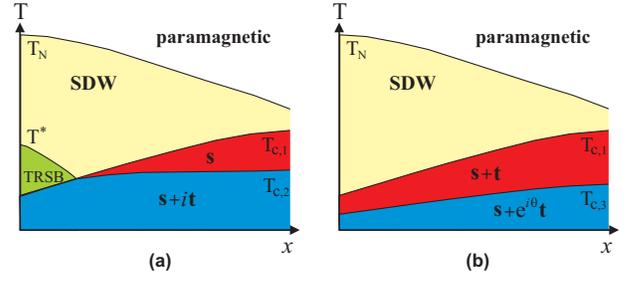}
\caption{
Schematic phase diagram of a superconducor in coexistence with SDW.
(a) The special case when $s$ and $t$ order parameters do not couple
linearly (nested FSs). (b) The generic case when $s$ and $t$ superconducting
components couple linearly (non-nested FSs). While in the $s-$phase
superconductivity has only a singlet component, in the $s+t$ phase
both singlet and triplet components are present but TRS is not broken.
In the $s+e^{i}\theta t$ and $s+it$ phases ($\theta=\pi/2)$, the
relative phase between the $s$ and $t$ components is frozen at $0<\theta<\pi$
and TRS is broken, together with the $U(1)$ symmetry of the global
phase. In the TRSB phase, only TRS is broken. This phase is likely present
in the generic case but its boundaries are not known and we do not show it. \label{fig:phase_dia}}
\end{figure}

To understand the physics, we first focus on the case of ``maximally-nested''
FSs, where $\delta_{0}=0$ but $\delta_{2}\neq0$, i.e. $\xi_{\mathbf{k}}^{b}$
becomes $-\xi_{\mathbf{k}}^{a}$ under a rotation by 90 degrees. We
found that this symmetry decouples the three linearized gap equations
for $\Delta_{i}$, which become
\begin{eqnarray}
 &  & \Delta_{1}\left[1-\frac{U_{3}-U_{4}}{2}N_{F}\int X_{\mathbf{k}}\right]=0\label{ch2}\\
 &  & \Delta_{2}\left[1-(U_{2}-U_{1})N_{F}\int\left(u_{\mathbf{k}}^{2}v_{\mathbf{k}}^{2}X_{\mathbf{k}}+(u_{\mathbf{k}}^{2}-v_{\mathbf{k}}^{2})^{2}Y_{\mathbf{k}}\right)\right]=0\nonumber \\
 &  & \Delta_{3}\left[1+\frac{U_{3}+U_{4}}{2}N_{F}\int\left((u_{\mathbf{k}}^{2}-v_{\mathbf{k}}^{2})^{2}X_{\mathbf{k}}+8u_{\mathbf{k}}^{2}v_{\mathbf{k}}^{2}Y_{\mathbf{k}}\right)\right]=0\nonumber
\end{eqnarray}
 where $N_{F}$ is the density of states at the FS, $\int=\int d\xi\frac{d\phi}{2\pi}$,
$u_{\mathbf{k}}v_{\mathbf{k}}=M/(2\sqrt{M^{2}+\xi_{\mathbf{k}}^{2}})$
and $u_{\mathbf{k}}^{2}-v_{\mathbf{k}}^{2}=\xi_{\mathbf{k}}/\sqrt{M^{2}+\xi_{\mathbf{k}}^{2}}$,
and
\begin{equation}
X_{\mathbf{k}}=\frac{\tanh{\frac{\xi_{\mathbf{k}}^{a}}{2T_{c}}}}{\xi_{\mathbf{k}}^{a}},~~Y_{\mathbf{k}}=\frac{\tanh{\frac{\xi_{\mathbf{k}}^{a}}{2T_{c}}}+\tanh{\frac{\xi_{\mathbf{k}}^{b}}{2T_{c}}}}{2(\xi_{\mathbf{k}}^{a}+\xi_{\mathbf{k}}^{b})}\label{ch5}
\end{equation}

The first and the last Eqs. (\ref{ch2}) have familiar forms for $s^{+-}$
and $s^{++}$ superconductivity, respectively~\cite{comm}. For positive
$U_{i}$, the $s^{++}$ channel is repulsive, but $s^{+-}$ superconductivity
develops at $T=T_{c,1}$ if $U_{3}-U_{4}$ is positive. The momentum
integral $\int X_{\mathbf{k}}$ is logarithmically singular, as expected
in BCS theory, hence $T_{c,1}$ is non-zero already at weak coupling.
The second Eq. in (\ref{ch2}) is the gap equation in the new pairing
channel. In the presence of SDW the kernel in this channel is also
logarithmically singular due to
the contribution from $\langle aa\rangle_{\mathbf{k}}+\langle bb\rangle_{\mathbf{k}}$.
Hence, if $U_{2}-U_{1}$ is positive, the $t-$channel becomes unstable
towards pairing at a non-zero $T_{c,2}$. Once $\Delta_{2}$ becomes
non-zero, it induces a non-zero inter-pocket pairing component $\langle ab\rangle_{\mathbf{k}}$,
which, due to the folding of the Brillouin zone imposed by SDW order,
$\mathbf{k}+\mathbf{Q}\rightarrow\mathbf{k}$, has zero center-of-mass
momentum.

\textit{$s+it$ state with broken time-reversal symmetry}~~~ As
it is customary for competing SC orders, the order which develops
first tends to suppress the competitor by providing negative feedback
to the gap equation for the competing order \cite{s_plus_is}. Yet,
if the repulsion between the competing SC orders is not too strong,
the two orders coexist at low enough temperatures. The issue then
is what is the relative phase between the two $U(1)$ order parameters
$\Delta_{1}$ and $\Delta_{2}$. To address this issue we derived
by standard means~\cite{Fernandes12,Nandkishore12} the GL Free energy,
${\cal F}(\Delta_{1},\Delta_{2})$ (see SM). To fourth order in $\Delta_{1,2}$
we obtained
\begin{align}
{\cal F}(\Delta_{1},\Delta_{2}) & =\alpha_{1}|\Delta_{1}|^{2}+\alpha_{2}|\Delta_{2}|^{2}+\beta_{1}|\Delta_{1}|^{4}+\beta_{2}|\Delta_{2}|^{4}\nonumber \\
 & +2\gamma_{1}|\Delta_{1}|^{2}|\Delta_{2}|^{2}+\gamma_{2}\left(\Delta_{1}^{2}(\Delta_{2}^{*})^{2}+(\Delta_{1}^{*})^{2}\Delta_{2}^{2}\right)\label{eq:deltaf}
\end{align}
 where $\beta_{1}$ and $\beta_{2}$ are positive. The two orders
coexist when $\beta_{1}\beta_{2}>\left(\gamma_{1}-\left|\gamma_{2}\right|\right)^{2}$.
This condition can be satisfied in the presence of disorder \cite{Vavilov11,Schmalian14}.
The relative phase $\theta$ between $\Delta_{1}=|\Delta_{1}|e^{i\psi+\theta/2}$
and $\Delta_{2}=|\Delta_{2}|e^{i\psi-\theta/2}$ is determined by
the sign of the $\gamma_{2}$ term in (\ref{eq:deltaf}). We found
that $\gamma_{2}$ is positive:
\begin{equation}
\gamma_{2}=\sum_{\mathbf{k}}(2u_{\mathbf{k}}v_{\mathbf{k}})^{2}\left[\frac{1}{|\xi_{\mathbf{k}}^{a}|^{3}}+\frac{1}{|\xi_{\mathbf{k}}^{b}|^{3}}\right].
\end{equation}
 Minimization of Eq. (\ref{eq:deltaf}) then shows that $\theta=\pm\pi/2$.
Because $\theta=\pi/2$ and $\theta=-\pi/2$ are different states,
the system spontaneously breaks the $Z_{2}$ TRS. In the TRS-broken
state, the phases of the order parameters $\langle aa\rangle_{\mathbf{k}}$
and $\langle bb\rangle_{\mathbf{k}}$ are $\phi$ and $\pi-\phi$,
where $0<\phi<\pi/2$. The third gap, which is generally required
to satisfy the set of complex gap equations in TRS-broken state is
provided by $\langle ab\rangle_{\mathbf{k}}$, whose phase in this
situation is
 $-\pi/2$. We show the gap structure schematically in Fig. \ref{fig:phasors}
where we associated $\langle ij\rangle_{\mathbf{k}}$ with vectors,
whose directions
 are set by the phases. We also performed Hubbard-Stratonovich analysis
beyond mean-field level \cite{Fernandes12}, by allowing the phases
of $\Delta_{1,2}$ to fluctuate, and found (see SM) that when $T_{c,2}\approx T_{c,1}\equiv T_{c}$,
the system breaks TRS and sets the relative phase
 $\theta=\pm\pi/2$ at a temperature $T^{*}>T_{c}$
. In between $T^{*}$ and $T_{c}$, TRS is broken, but the $U(1)$
symmetry associated with the global phase of $\Delta_{1}$ and $\Delta_{2}$
remains intact. At $T_{c}$, the global phase is broken and both SC
orders develop simultaneously.
A schematic phase diagram is shown in Fig. \ref{fig:phase_dia}a.

\textit{$s+\mathrm{e}^{i\theta}t$ state}~~~ So far we considered
the ``maximally-nested'' case, with $\delta_{0}=0$. For
the more generic case $\delta_{0}\neq0$ we find that the GL functional
(\ref{eq:deltaf}) contains a bilinear coupling between the two SC
states, i.e. a term $\alpha_{3}\left(\Delta_{1}\Delta_{2}^{*}+\Delta_{1}^{*}\Delta_{2}\right)$
with $\alpha_{3}<0$ (details in the SM). In this situation, the onset
of the $s^{+-}$ state at $T_{c,1}$ necessarily triggers the emergence
of a $t$ state. The relative phase between the two order parameters
at $T\leq T_{c,1}$ is $\theta=0$, i.e., the state is $s+t$. Yet,
the SC state still breaks TRS at a lower temperature $T_{c,3}<T_{c,1}$.
Indeed, comparing the $\alpha_{3}\left(\Delta_{1}\Delta_{2}^{*}+\Delta_{1}^{*}\Delta_{2}\right)$
and $\gamma_{2}\left(\Delta_{1}^{2}(\Delta_{2}^{*})^{2}+(\Delta_{1}^{*})^{2}\Delta_{2}^{2}\right)$
terms in the GL functional we immediately see that $\theta=0$ only
as long as $\Delta_{1}\Delta_{2}<\alpha_{3}/4\gamma_{2}$. Once the
temperature is reduced and $\Delta_{1,2}$ grow, this condition breaks
down at $T=T_{c,3}$, and at lower $T$ the minimum of
the GL functional shifts to $\theta\neq0$. Once this happens, the
SC state becomes $s+\mathrm{e}^{i\theta}t$ and TRS gets broken.
A schematic phase diagram is shown in Fig. \ref{fig:phase_dia}b.

\textit{Conclusions}~~~ In this paper we argued that a SC state,
which explicitly breaks TRS, appears when SC emerges from a pre-existing
SDW-odered state. We found that in the presence of SDW, the spin-triplet
channel with inter-pocket pairing couples to spin-singlet intra-pocket
pairings on the reconstructed FSs. This leads to the emergence of
a new pairing channel, which we labeled as $t-$pairing to emphasize
that it involves spin-triplet. We analyzed the interplay between $s^{+-}$
and $t-$ SC orders and showed that they coexist at low $T$ with
a relative phase $0<\theta<\pi$. As a result, the phases of the gaps
on different FSs differ by less
than a multiple of $\pi$. Such a state breaks time-reversal symmetry
and has been long south in the studies of FeSCs. We argued that in
a generic case TRS gets broken in the SC manifold at temperatures
lower than $T_{c}$. This should give rise to features in experimentally
probed thermodynamic quantities.

We thank P. Hirschfeld, I. Eremin, and O. Vafek for fruitful discussions.
AVC and AH are supported by the DOE grant DE-FG02-ER46900.

\section{Supplementary Material}~~~

In the Supplementary Material we
discuss some technical details of the analysis presented in the
main text.

\subsection{Interaction Hamiltonian}~~~ We include all five possible repulsive interactions in the band basis
\begin{align}
 & \mathcal{H}_{int}=U_{1}\sum c_{\mathbf{p}_{3}\sigma}^{\dagger}f_{\mathbf{p}_{4}\sigma^{\prime}}^{\dagger}f_{\mathbf{p}_{2}\sigma^{\prime}}c_{\mathbf{p}_{1}\sigma},\nonumber \\
 & +U_{2}\sum f_{\mathbf{p}_{3}\sigma}^{\dagger}c_{\mathbf{p}_{4}\sigma^{\prime}}^{\dagger}f_{\mathbf{p}_{2}\sigma^{\prime}}c_{\mathbf{p}_{1}\sigma},\nonumber \\
 & +\frac{U_{3}}{2}\sum\left[f_{\mathbf{p}_{3}\sigma}^{\dagger}f_{\mathbf{p}_{4}\sigma^{\prime}}^{\dagger}c_{\mathbf{p}_{2}\sigma^{\prime}}c_{\mathbf{p}_{1}\sigma}+\mathrm{H.c.}\right],\label{eq:interactions}\\
 & +\frac{U_{4}}{2}\sum f_{\mathbf{p}_{3}\sigma}^{\dagger}f_{\mathbf{p}_{4}\sigma^{\prime}}^{\dagger}f_{\mathbf{p}_{2}\sigma^{\prime}}f_{\mathbf{p}_{1}\sigma}+\frac{U_{5}}{2}\sum c_{\mathbf{p}_{3}\sigma}^{\dagger}c_{\mathbf{p}_{4}\sigma^{\prime}}^{\dagger}c_{\mathbf{p}_{2}\sigma^{\prime}}c_{\mathbf{p}_{1}\sigma}.\nonumber
\end{align}
 The momentum conservation is implicit and $\sigma\neq\sigma^{\prime}$
in all sums. The first three are inter-pocket density-density, exchange,
and pair hopping, interactions, respectively (all positive), while
the last two are intra-pocket repulsions. For simplicity, we set $U_{4}=U_{5}$
below. All couplings are assumed to be already renormalized from their
bare values by fermions with energies larger than the upper energy
cutoff $\Lambda$.

\subsection{SDW state}~~~ In order to introduce the SDW order starting from the paramagnetic state we first write the quadratic part of the Hamiltonian in the mean-field approximation, where the order parameter $M$ is defined as
\begin{align}
	M &=-\frac{U_1+U_3}{2}\sum_\mathbf{p} \sigma^z_{\alpha\beta} \left\langle c^\dagger_{\mathbf{p}\alpha} f_{\mathbf{p}\beta}\right\rangle,\nonumber\\
	 &=-\frac{U_1+U_3}{2}\sum_\mathbf{p} \sigma^z_{\alpha\beta} \left\langle f^\dagger_{\mathbf{p}\alpha} c_{\mathbf{p}\beta}\right\rangle.
\end{align}
Then we perform the following Bogoliubov transformation to bring it to diagonal form:
\begin{align}
	c_{\mathbf{k}\alpha} &= u_\mathbf{k} a_{\mathbf{k}\alpha}+v_\mathbf{k} \sigma^z_{\alpha\beta} b_{\mathbf{k}\beta},\label{eq:bogoliubov1},\\
	f_{\mathbf{k}\alpha} &= u_\mathbf{k} b_{\mathbf{k}\alpha}-v_\mathbf{k} \sigma^z_{\alpha\beta} a_{\mathbf{k}\beta}.\label{eq:bogoliubov2}.
\end{align}

\subsection{Gap equations}~~~ In order to carry out the diagonalization
of the mean-field Hamiltonian (\ref{eq:mfH}) we apply the following
Bogoliubov transformation, introducing new quasiparticle operators
$\alpha$ and $\beta$:
\begin{align}
a_{k\mu} & =u_{\mathbf{k}}^{\alpha}\alpha_{\mathbf{k}\mu}+v_{\mathbf{k}}^{\alpha}i\sigma_{\mu\nu}^{y}\alpha_{-\mathbf{k}\nu}^{\dagger}+g_{\mathbf{k}}^{\beta}\sigma_{\mu\nu}^{z}\beta_{\mathbf{k}\nu}+h_{\mathbf{k}}^{\beta}\sigma_{\mu\nu}^{x}\beta_{-\mathbf{k}\nu}^{\dagger},\\
b_{k\mu} & =u_{\mathbf{k}}^{\beta}\beta_{\mathbf{k}\mu}+v_{\mathbf{k}}^{\beta}i\sigma_{\mu\nu}^{y}\beta_{-\mathbf{k}\nu}^{\dagger}+g_{\mathbf{k}}^{\alpha}\sigma_{\mu\nu}^{z}\alpha_{\mathbf{k}\nu}+h_{\mathbf{k}}^{\alpha}\sigma_{\mu\nu}^{x}\alpha_{-\mathbf{k}\nu}^{\dagger}.
\end{align}
 As a result, we obtain a quadratic Hamiltonian
\begin{equation}
\mathcal{H}=\sum_{\mathbf{k},\mu}\left[E_{\mathbf{k}}^{\alpha}\alpha_{\mathbf{k}\mu}^{\dagger}\alpha_{\mathbf{k}\mu}+E_{\mathbf{k}}^{\beta}\beta_{\mathbf{k}\mu}^{\dagger}\beta_{\mathbf{k}\mu}\right]
\end{equation}
 and new quasiparticle dispersions
\begin{equation}
E_{\mathbf{k}}^{\alpha,\beta}=\sqrt{A_{\mathbf{k}}\pm\sqrt{B_{\mathbf{k}}}},
\end{equation}
 where
\begin{equation}
A_{\mathbf{k}}=\frac{(\xi_{\mathbf{k}}^{\alpha})^{2}+(\xi_{\mathbf{k}}^{\beta})^{2}}{2}+|\Delta_{1}|^{2}+|\Delta_{2}|^{2}+|\Delta_{3}|^{2},
\end{equation}

\begin{align}
B_{\mathbf{k}} & =\left(\frac{(\xi_{\mathbf{k}}^{\alpha})^{2}-(\xi_{\mathbf{k}}^{\beta})^{2}}{2}\right)^{2}\nonumber \\
 & +\left[(\xi_{\mathbf{k}}^{\alpha})^{2}-(\xi_{\mathbf{k}}^{\beta})^{2}\right][t(\Delta_{1}\Delta_{3}^{*}+\Delta_{1}^{*}\Delta_{3})+s(\Delta_{1}\Delta_{2}^{*}+\Delta_{1}^{*}\Delta_{2})]\nonumber \\
 & +\left[\xi_{\mathbf{k}}^{\alpha}-\xi_{\mathbf{k}}^{\beta}\right]^{2}(t\Delta_{2}-s\Delta_{3})(t\Delta_{2}^{*}-s\Delta_{3}^{*})\nonumber \\
 & +(\Delta_{1}\Delta_{2}^{*}+\Delta_{1}^{*}\Delta_{2})^{2}+(\Delta_{1}\Delta_{3}^{*}+\Delta_{1}^{*}\Delta_{3})^{2}\nonumber \\
 & -(\Delta_{2}\Delta_{3}^{*}-\Delta_{2}^{*}\Delta_{3})^{2},
\end{align}
 and we have defined $s\equiv\frac{M}{\sqrt{M^{2}+\xi_{\mathbf{k}}^{2}}}$
and $t\equiv\frac{\xi_{\mathbf{k}}}{\sqrt{M^{2}+\xi_{\mathbf{k}}^{2}}}$.

The gap equations can be found by starting with the expressions for
the order parameters,
\begin{align}
\Delta_{1} & =\frac{U_{3}-U_{4}}{2}\sum_{\mathbf{k}}\left[\langle aa\rangle_{\mathbf{k}}-\langle bb\rangle_{\mathbf{k}}\right],\\
\Delta_{2} & =(U_{2}-U_{1})\sum_{\mathbf{k}}\left[u_{\mathbf{k}}v_{\mathbf{k}}(\langle aa\rangle_{\mathbf{k}}+\langle bb\rangle_{\mathbf{k}})+(u_{\mathbf{k}}^{2}-v_{\mathbf{k}}^{2})\langle ab\rangle_{\mathbf{k}}\right],\\
\Delta_{3} & =-\frac{U_{3}+U_{4}}{2}\sum_{\mathbf{k}}\left[(u_{\mathbf{k}}^{2}-v_{\mathbf{k}}^{2})(\langle aa\rangle_{\mathbf{k}}+\langle bb\rangle_{\mathbf{k}})-4u_{\mathbf{k}}v_{\mathbf{k}}\langle ab\rangle_{\mathbf{k}}\right],
\end{align}
 and substituting the following expressions for the averages $\langle ij\rangle$:
\begin{align}
\langle aa\rangle_{\mathbf{k}}-\langle bb\rangle_{\mathbf{k}} & =-(u_{\mathbf{k}}^{\alpha}v_{\mathbf{k}}^{\alpha}+g_{\mathbf{k}}^{\alpha}h_{\mathbf{k}}^{\alpha})(1-2n_{F}(E_{\mathbf{k}}^{\alpha}))\nonumber \\
 & +(u_{\mathbf{k}}^{\beta}v_{\mathbf{k}}^{\beta}+g_{\mathbf{k}}^{\beta}h_{\mathbf{k}}^{\beta})(1-2n_{F}(E_{\mathbf{k}}^{\beta})),\\
\langle aa\rangle_{\mathbf{k}}+\langle bb\rangle_{\mathbf{k}} & =(-u_{\mathbf{k}}^{\alpha}v_{\mathbf{k}}^{\alpha}+g_{\mathbf{k}}^{\alpha}h_{\mathbf{k}}^{\alpha})(1-2n_{F}(E_{\mathbf{k}}^{\alpha}))\nonumber \\
 & +(-u_{\mathbf{k}}^{\beta}v_{\mathbf{k}}^{\beta}+g_{\mathbf{k}}^{\beta}h_{\mathbf{k}}^{\beta})(1-2n_{F}(E_{\mathbf{k}}^{\beta})),\\
\langle ab\rangle_{\mathbf{k}} & =u_{\mathbf{k}}^{\alpha}h_{\mathbf{k}}^{\alpha}(1-n_{F}(E_{\mathbf{k}}^{\alpha}))-v_{\mathbf{k}}^{\alpha}g_{\mathbf{k}}^{\alpha}n_{F}(E_{\mathbf{k}}^{\alpha})\nonumber \\
 & +u_{\mathbf{k}}^{\beta}h_{\mathbf{k}}^{\beta}n_{F}(E_{\mathbf{k}}^{\beta})-v_{\mathbf{k}}^{\beta}g_{\mathbf{k}}^{\beta}(1-n_{F}(E_{\mathbf{k}}^{\beta})),
\end{align}
 where $n_{F}$ is the Fermi distribution function.

The coherence factors are given by
\begin{align}
\left(u_{\mathbf{k}}^{\alpha},v_{\mathbf{k}}^{\alpha},g_{\mathbf{k}}^{\alpha},h_{\mathbf{k}}^{\alpha}\right) & =\frac{\left(U_{\mathbf{k}}^{\alpha},V_{\mathbf{k}}^{\alpha},G_{\mathbf{k}}^{\alpha},H_{\mathbf{k}}^{\alpha}\right)}{\sqrt{|U_{\mathbf{k}}^{\alpha}|^{2}+|V_{\mathbf{k}}^{\alpha}|^{2}+|G_{\mathbf{k}}^{\alpha}|^{2}+|H_{\mathbf{k}}^{\alpha}|^{2}}}\\
\left(u_{\mathbf{k}}^{\beta},v_{\mathbf{k}}^{\beta},g_{\mathbf{k}}^{\beta},h_{\mathbf{k}}^{\beta}\right) & =\frac{\left(U_{\mathbf{k}}^{\beta},V_{\mathbf{k}}^{\beta},G_{\mathbf{k}}^{\beta},H_{\mathbf{k}}^{\beta}\right)}{\sqrt{|U_{\mathbf{k}}^{\beta}|^{2}+|V_{\mathbf{k}}^{\beta}|^{2}+|G_{\mathbf{k}}^{\beta}|^{2}+|H_{\mathbf{k}}^{\beta}|^{2}}},
\end{align}
 where
\begin{align}
U_{\mathbf{k}}^{\alpha} & =[E_{\mathbf{k}}^{\alpha}+\xi_{\mathbf{k}}^{a}][-|\Delta_{1}|^{2}-|\Delta_{2}|^{2}+s(\Delta_{1}\Delta_{2}^{*}+\Delta_{1}^{*}\Delta_{2})\nonumber \\
 & \quad\quad+t(\Delta_{1}\Delta_{3}^{*}+\Delta_{1}^{*}\Delta_{3})]\nonumber \\
 & +t[\xi_{\mathbf{k}}^{a}-\xi_{\mathbf{k}}^{b}][t|\Delta_{2}|^{2}-t|\Delta_{3}|^{2}+s(\Delta_{2}\Delta_{3}^{*}+\Delta_{2}^{*}\Delta_{3})]\nonumber \\
 & +[E_{\mathbf{k}}^{\alpha}+\xi_{\mathbf{k}}^{b}][(E_{\mathbf{k}}^{\alpha}+\xi_{\mathbf{k}}^{a})(E_{\mathbf{k}}^{\alpha}-\xi_{\mathbf{k}}^{b})-|\Delta_{3}|^{2}]
\end{align}

\begin{align}
U_{\mathbf{k}}^{\beta} & =[E_{\mathbf{k}}^{\beta}+\xi_{\mathbf{k}}^{b}][-|\Delta_{1}|^{2}-|\Delta_{2}|^{2}-s(\Delta_{1}\Delta_{2}^{*}+\Delta_{1}^{*}\Delta_{2})\nonumber \\
 & \quad\quad-t(\Delta_{1}\Delta_{3}^{*}+\Delta_{1}^{*}\Delta_{3})]\nonumber \\
 & +t[\xi_{\mathbf{k}}^{b}-\xi_{\mathbf{k}}^{a}][t|\Delta_{2}|^{2}-t|\Delta_{3}|^{2}+s(\Delta_{2}\Delta_{3}^{*}+\Delta_{2}^{*}\Delta_{3})]\nonumber \\
 & +[E_{\mathbf{k}}^{\beta}+\xi_{\mathbf{k}}^{a}][(E_{\mathbf{k}}^{\beta}+\xi_{\mathbf{k}}^{b})(E_{\mathbf{k}}^{\beta}-\xi_{\mathbf{k}}^{a})-|\Delta_{3}|^{2}]
\end{align}

\begin{align}
V_{\mathbf{k}}^{\alpha} & =[-(E_{\mathbf{k}}^{\alpha})^{2}+(\xi_{\mathbf{k}}^{b})^{2}][\Delta_{1}+s\Delta_{2}+t\Delta_{3}]\nonumber \\
 & +[\Delta_{1}^{2}-\Delta_{2}^{2}-\Delta_{3}^{2}][\Delta_{1}^{*}-s\Delta_{2}^{*}-t\Delta_{3}^{*}]
\end{align}

\begin{align}
V_{\mathbf{k}}^{\beta} & =[-(E_{\mathbf{k}}^{\beta})^{2}+(\xi_{\mathbf{k}}^{a})^{2}][-\Delta_{1}+s\Delta_{2}+t\Delta_{3}]\nonumber \\
 & +[\Delta_{1}^{2}-\Delta_{2}^{2}-\Delta_{3}^{2}][-\Delta_{1}^{*}-s\Delta_{2}^{*}-t\Delta_{3}^{*}]
\end{align}

\begin{align}
G_{\mathbf{k}}^{\alpha} & =[E_{\mathbf{k}}^{\alpha}+\xi_{\mathbf{k}}^{a}][t\Delta_{1}\Delta_{2}^{*}-s\Delta_{1}\Delta_{3}^{*}+\Delta_{2}\Delta_{3}^{*}]\nonumber \\
 & +[E_{\mathbf{k}}^{\alpha}+\xi_{\mathbf{k}}^{b}][t\Delta_{1}^{*}\Delta_{2}-s\Delta_{1}^{*}\Delta_{3}-\Delta_{2}^{*}\Delta_{3}]\nonumber \\
 & +t[\xi_{\mathbf{k}}^{a}-\xi_{\mathbf{k}}^{b}][s(-|\Delta_{2}|^{2}+|\Delta_{3}|^{2})-t(\Delta_{2}\Delta_{3}^{*}+\Delta_{2}^{*}\Delta_{3})]
\end{align}

\begin{align}
G_{\mathbf{k}}^{\beta} & =[E_{\mathbf{k}}^{\beta}+\xi_{\mathbf{k}}^{b}][t\Delta_{1}\Delta_{2}^{*}-s\Delta_{1}\Delta_{3}^{*}-\Delta_{2}\Delta_{3}^{*}]\nonumber \\
 & +[E_{\mathbf{k}}^{\beta}+\xi_{\mathbf{k}}^{a}][t\Delta_{1}^{*}\Delta_{2}-s\Delta_{1}^{*}\Delta_{3}+\Delta_{2}^{*}\Delta_{3}]\nonumber \\
 & +t[\xi_{\mathbf{k}}^{a}-\xi_{\mathbf{k}}^{b}][s(-|\Delta_{2}|^{2}+|\Delta_{3}|^{2})-t(\Delta_{2}\Delta_{3}^{*}+\Delta_{2}^{*}\Delta_{3})]
\end{align}

\begin{align}
H_{\mathbf{k}}^{\alpha} & =[E_{\mathbf{k}}^{\alpha}+\xi_{\mathbf{k}}^{a}][E_{\mathbf{k}}^{\alpha}-\xi_{\mathbf{k}}^{b}][t\Delta_{2}-s\Delta_{3}]\nonumber \\
 & +[\Delta_{1}^{2}-\Delta_{2}^{2}-\Delta_{3}^{2}][t\Delta_{2}^{*}-s\Delta_{3}^{*}]
\end{align}

\begin{align}
H_{\mathbf{k}}^{\beta} & =[E_{\mathbf{k}}^{\beta}+\xi_{\mathbf{k}}^{b}][E_{\mathbf{k}}^{\beta}-\xi_{\mathbf{k}}^{a}][-t\Delta_{2}+s\Delta_{3}]\nonumber \\
 & +[\Delta_{1}^{2}-\Delta_{2}^{2}-\Delta_{3}^{2}][-t\Delta_{2}^{*}+s\Delta_{3}^{*}]
\end{align}

The expansion of the gap equations to linear order in $\Delta_{i}$
yields

\begin{align}
\Delta_{1} & =\frac{U_{3}-U_{4}}{2}\sum_{\mathbf{k}}\Bigg\{\Delta_{1}\left[\frac{\tanh(\xi_{\mathbf{k}}^{a}/(2T))}{2\xi_{\mathbf{k}}^{a}}+(a\rightarrow b)\right]\Bigg.\nonumber \\
 & +\Delta_{2}\Bigg.s\left[\frac{\tanh(\xi_{\mathbf{k}}^{a}/(2T))}{2\xi_{\mathbf{k}}^{a}}-(a\rightarrow b)\Bigg]\right\}
\end{align}

\begin{align}
\Delta_{2} & =(U_{2}-U_{1})\sum_{\mathbf{k}}\Bigg\{\Delta_{2}\frac{s^{2}}{2}\left[\frac{\tanh(\xi_{\mathbf{k}}^{a}/(2T))}{2\xi_{\mathbf{k}}^{a}}+(a\rightarrow b)\right]\Bigg.\nonumber \\
 & +\Delta_{2}t^{2}\left[\frac{\tanh(\xi_{\mathbf{k}}^{a}/(2T))+\tanh(\xi_{\mathbf{k}}^{b}/(2T))}{2(\xi_{\mathbf{k}}^{a}+\xi_{\mathbf{k}}^{b})}\right]\nonumber \\
 & +\Delta_{1}\Bigg.\frac{s}{2}\left[\frac{\tanh(\xi_{\mathbf{k}}^{a}/(2T))}{2\xi_{\mathbf{k}}^{a}}-(a\rightarrow b)\Bigg]\right\}
\end{align}

\begin{align}
\Delta_{3} & =-\frac{U_{3}+U_{4}}{2}\Delta_{3}\sum_{\mathbf{k}}\Bigg\{ t^{2}\left[\frac{\tanh(\xi_{\mathbf{k}}^{a}/(2T))}{2\xi_{\mathbf{k}}^{a}}+(a\rightarrow b)\right]\Bigg.\nonumber \\
 & +\Bigg.2s^{2}\left[\frac{\tanh(\xi_{\mathbf{k}}^{a}/(2T))+\tanh(\xi_{\mathbf{k}}^{b}/(2T))}{2(\xi_{\mathbf{k}}^{a}+\xi_{\mathbf{k}}^{b})}\right]\Bigg\}
\end{align}

\subsection{Coexistence of superconducting orders}~~~ We present the
conditions that are necessary for the coexistence of the $\Delta_{1}$
and $\Delta_{2}$ orders. We begin by listing the full expressions
for all the coefficients of the free energy.

\begin{align}
{\cal F}(\Delta_{1},\Delta_{2}) & =\alpha_{1}|\Delta_{1}|^{2}+\alpha_{2}|\Delta_{2}|^{2}+\alpha_{3}(\Delta_{1}\Delta_{2}^{*}+\Delta_{1}^{*}\Delta_{2})\nonumber \\
 & +\beta_{1}|\Delta_{1}|^{4}+\beta_{2}|\Delta_{2}|^{4}+2\gamma_{1}|\Delta_{1}|^{2}|\Delta_{2}|^{2}\nonumber \\
 & +\gamma_{2}\left(\Delta_{1}^{2}(\Delta_{2}^{*})^{2}+(\Delta_{1}^{*})^{2}\Delta_{2}^{2}\right)
\end{align}

\begin{align}
\alpha_{1} & =-\frac{1}{2}\sum_{\mathbf{k}}\left[\frac{1}{|\xi_{\mathbf{k}}^{a}|}+\frac{1}{|\xi_{\mathbf{k}}^{b}|}\right]+\frac{2}{U_{3}-U_{4}},\\
\alpha_{2} & =-\frac{1}{2}\sum_{\mathbf{k}}s^{2}\left[\frac{1}{|\xi_{\mathbf{k}}^{a}|}+\frac{1}{|\xi_{\mathbf{k}}^{b}|}\right]\nonumber \\
 & -\sum_{\mathbf{k}}t^{2}\frac{\sgn\xi_{\mathbf{k}}^{a}+\sgn\xi_{\mathbf{k}}^{b}}{\xi_{\mathbf{k}}^{a}+\xi_{\mathbf{k}}^{b}}+\frac{2}{U_{2}-U_{1}},\\
\alpha_{3} & =-\frac{1}{2}\sum_{\mathbf{k}}s\left[\frac{1}{|\xi_{\mathbf{k}}^{a}|}-\frac{1}{|\xi_{\mathbf{k}}^{b}|}\right],\\
\beta_{1} & =\frac{1}{8}\sum_{\mathbf{k}}\left[\frac{1}{|\xi_{\mathbf{k}}^{a}|^{3}}+\frac{1}{|\xi_{\mathbf{k}}^{b}|^{3}}\right],\\
\beta_{2} & =\frac{1}{8}\sum_{\mathbf{k}}s^{4}\left[\frac{1}{|\xi_{\mathbf{k}}^{a}|^{3}}+\frac{1}{|\xi_{\mathbf{k}}^{b}|^{3}}\right]\nonumber \\
 & +\sum_{\mathbf{k}}t^{4}\left[\frac{\sgn\xi_{\mathbf{k}}^{a}+\sgn\xi_{\mathbf{k}}^{b}}{(\xi_{\mathbf{k}}^{a}+\xi_{\mathbf{k}}^{b})^{3}}\right]\\
\gamma_{1} & =\gamma_{2}+\frac{1}{8}\sum_{\mathbf{k}}s^{2}\left[\frac{1}{|\xi_{\mathbf{k}}^{a}|^{3}}+\frac{1}{|\xi_{\mathbf{k}}^{b}|^{3}}\right]\nonumber \\
 & +\frac{1}{4}\sum_{\mathbf{k}}t^{2}\left[\frac{\sgn\xi_{\mathbf{k}}^{a}}{(\xi_{\mathbf{k}}^{a})^{2}(\xi_{\mathbf{k}}^{a}+\xi_{\mathbf{k}}^{b})}+\frac{\sgn\xi_{\mathbf{k}}^{b}}{(\xi_{\mathbf{k}}^{b})^{2}(\xi_{\mathbf{k}}^{a}+\xi_{\mathbf{k}}^{b})}\right],\\
\gamma_{2} & =\frac{1}{8}\sum_{\mathbf{k}}s^{2}\left[\frac{1}{|\xi_{\mathbf{k}}^{a}|^{3}}+\frac{1}{|\xi_{\mathbf{k}}^{b}|^{3}}\right]\nonumber \\
 & +\frac{1}{4}\sum_{\mathbf{k}}\frac{t^{2}}{(\xi_{\mathbf{k}}^{a})^{2}-(\xi_{\mathbf{k}}^{b})^{2}}\left[-\frac{1}{|\xi_{\mathbf{k}}^{a}|}+\frac{1}{|\xi_{\mathbf{k}}^{b}|}\right].
\end{align}

The largest contribution to these integrals comes from the regions
around $\xi_{\mathbf{k}}^{a}=0$ and $\xi_{\mathbf{k}}^{b}=0$ (the
SDW FSs), where the denominators become zero. This singularity is
caused by calculating the coefficients at $T=0$ and is removed by
including a small cutoff at those points. One may think that the regions
where $\xi_{\mathbf{k}}^{a}+\xi_{\mathbf{k}}^{a}=0$ are also singular
but in each case the integrand is actually finite. Thus the main contributions
to the coefficients $\beta_{i}$ and $\gamma_{i}$ are the integrals
with $|\xi_{\mathbf{k}}^{a,b}|^{-3}$. All of these are positive definite
so $\beta_{i}>0$ and $\gamma_{i}>0$.

In the case of $\delta_{0}=0$ the coefficient $\alpha_{3}$ vanishes,
so the order parameters decouple at linear order. To determine whether
coexistance occurs we search for minima of the free energy where both
parameters are non-zero. First note that the remaining terms depend
only on $|\Delta_{1}|^{2}$ and $|\Delta_{2}|^{2}$, except for the
term with coefficient $\gamma_{2}$. Since $\gamma_{2}>0$, the minimum
value of this term is $-2\gamma_{2}|\Delta_{1}|^{2}|\Delta_{2}|^{2}$,
which corresponds to a phase difference between $\Delta_{1}$ and
$\Delta_{2}$ of $\pm\pi/2$. After we fix this phase, partial differentiation
with respect to $|\Delta_{1}|^{2}$ and $|\Delta_{2}|^{2}$ yields
the following critical points:
\begin{align}
|\Delta_{1}|^{2}=\frac{\alpha_{2}(\gamma_{1}-\gamma_{2})-\alpha_{1}\beta_{2}}{2(\beta_{1}\beta_{2}-(\gamma_{1}-\gamma_{2})^{2})},\\
|\Delta_{2}|^{2}=\frac{\alpha_{1}(\gamma_{1}-\gamma_{2})-\alpha_{2}\beta_{1}}{2(\beta_{1}\beta_{2}-(\gamma_{1}-\gamma_{2})^{2})}.
\end{align}
 We then perform the second partial derivative test to find a necessary
condition for the existence of local minima. This condition is
\begin{equation}
\beta_{1}\beta_{2}>(\gamma_{1}-\gamma_{2})^{2}.
\end{equation}
 In addition, we require that the expressions for $|\Delta_{1}|^{2}$
and $|\Delta_{2}|^{2}$ be positive, which implies
\begin{align}
\alpha_{2}(\gamma_{1}-\gamma_{2})-\alpha_{1}\beta_{2} & >0,\\
\alpha_{1}(\gamma_{1}-\gamma_{2})-\alpha_{2}\beta_{1} & >0.
\end{align}
 Coexistence will occur if and only if all three inequalities are
satisfied.

\subsection{Preemtive TRS breaking above $T_{c}$}~~~ In this section
we show our Hubbard-Stratonovich analysis beyond mean-field level.
We take the Ginzburg-Landau free energy as an effective action and
study the case where $\delta_{0}=0$ and the critical temperatures
$T_{c1}\approx T_{c2}$. We consider an action of the form

\begin{align}
{\cal S}(\Delta_{1},\Delta_{2}) & =\alpha(|\Delta_{1}|^{2}+|\Delta_{2}|^{2})\nonumber \\
 & +\beta_{1}(|\Delta_{1}|^{2}+|\Delta_{2}|^{2})^{2}-\beta(|\Delta_{1}|^{2}-|\Delta_{2}|^{2})^{2}\nonumber \\
 & +\gamma\left(\Delta_{1}\Delta_{2}^{*}-\Delta_{1}^{*}\Delta_{2}\right)^{2}
\end{align}
 where $\alpha=a(T-T_{c})$ and $a$, $\beta_{1}$, $\beta$, and
$\gamma$ are positive. Then we apply a Hubbard-Stratonovich transformation
to this action by introducing collective variables $\tilde{\Phi}$,
$\Upsilon$, and $\Gamma$, which are conjugate to $(|\Delta_{1}|^{2}+|\Delta_{2}|^{2})^{2}$,
$(|\Delta_{1}|^{2}-|\Delta_{2}|^{2})^{2}$, and $\left(\Delta_{1}\Delta_{2}^{*}-\Delta_{1}^{*}\Delta_{2}\right)^{2}$,
respectively. By integrating out the fields $\Delta_{1}$ and $\Delta_{2}$
we obtain an effective action

\begin{align}
\mathcal{S}(\Phi,\Upsilon,\Gamma) & =\frac{\tilde{\Phi}^{2}}{4\beta_{1}}+\frac{\Upsilon^{2}}{4\beta}+\frac{\Gamma^{2}}{4\gamma}\\
 & +\int\frac{\mathrm{d}^{2}\mathbf{q}}{(2\pi)^{2}}\log\left[\left(\alpha-i\tilde{\Phi}+\mathbf{q}^{2}\right)^{2}-\Upsilon^{2}-\Gamma^{2}\right],\nonumber
\end{align}
 where we included the usual $\mathbf{q}^{2}$ dispersion in the quadratic
term by replacing $\alpha$ by $\alpha+\mathbf{q}^{2}$.

Now we seach for local minima of this action by differentiating with
respect to the three fields, obtaining a set of coupled equations.
The solution requires $\tilde{\Phi}$ to be purely imaginary, that
is $\tilde{\Phi}=i\Phi$. The set of equations becomes
\begin{align}
\Phi & =4\beta_{1}\int\frac{\mathrm{d}^{2}\mathbf{q}}{(2\pi)^{2}}\frac{\alpha+\Phi+\mathbf{q}^{2}}{\left(\alpha+\Phi+\mathbf{q}^{2}\right)^{2}-\Upsilon^{2}-\Gamma^{2}},\\
\Upsilon & =4\beta\int\frac{\mathrm{d}^{2}\mathbf{q}}{(2\pi)^{2}}\frac{\Upsilon}{\left(\alpha+\Phi+\mathbf{q}^{2}\right)^{2}-\Upsilon^{2}-\Gamma^{2}},\\
\Gamma & =4\gamma\int\frac{\mathrm{d}^{2}\mathbf{q}}{(2\pi)^{2}}\frac{\Gamma}{\left(\alpha+\Phi+\mathbf{q}^{2}\right)^{2}-\Upsilon^{2}-\Gamma^{2}}.
\end{align}
 Note that $\Gamma$ and $\Upsilon$ cannot simultaneously be nonzero
as a solution to these equations except in the special case of $\beta=\gamma$.

We first consider the solution with $\Gamma=\Upsilon=0$, which yields
\begin{equation}
\Phi=\frac{\beta_{1}}{\pi}\log\frac{\Lambda}{|\alpha+\Phi|},
\end{equation}
 where $\Lambda$ is an upper cutoff for the momentum integral. By
expanding the action about this solution we find that it is stable
as long as $\alpha>\mathrm{max}(\alpha_{cr1},\alpha_{cr2})$, where
\begin{align}
\alpha_{cr1}= & \frac{\gamma}{\pi}-\frac{\beta_{1}}{\pi}\log\frac{\pi\Lambda}{\gamma},\\
\alpha_{cr2}= & \frac{\beta}{\pi}-\frac{\beta_{1}}{\pi}\log\frac{\pi\Lambda}{\beta}
\end{align}
 This condition is equivalent to $T>T^*$ where $T^*=T_{c}+\mathrm{max}(\alpha_{cr1},\alpha_{cr2})/a$.
Whichever is greater between $\gamma$ and $\beta$ determines this
critical temperature. Then if $\gamma>\beta$ ($\gamma<\beta$) the
field $\Gamma$ ($\Upsilon$) will develop a nonzero solution and
the other one will remain zero. When we calculate $\beta$ and $\gamma$
in terms of the original coefficients of the Ginzburg-Landau free
energy we find that indeed $\gamma>\beta$. This means that a preemtive
order forms at a temperature above the critical temperature, where
time-reversal symmetry is broken before the gaps acquire non-zero
mean-field values.

This can be verified by solving the set of equations for $\Gamma\neq0$.
Expanding at small $\Gamma$ we find that
\begin{equation}
\Gamma^{2}\left(\frac{\beta_{1}}{\gamma}-2\right)\propto(T^*-T),
\end{equation}
 which means that if $\beta_{1}>2\gamma$ (which is satisfied in our
case) then $\Gamma$ gradually increases as $T$ becomes smaller than
$T^*$, as expected for a second-order transition.

\begin{thebibliography}{10}
\bibitem{comm} At a non-zero $\delta_{0}$, the gap in the $s^{+-}$
channel acquires some SDW-induced angular dependence due to ellipticity
of electron pockets~\cite{Maiti12_nodes}. This effect does not alter
our consideration.
\end{thebibliography}
\end{document}